\documentstyle[aps,prl,epsfig]{revtex}
\tightenlines
\newcommand{\be}{\begin{equation}}
\newcommand{\ee}{\end{equation}}
\newcommand{\ben}{\begin{eqnarray}}
\newcommand{\een}{\end{eqnarray}}
\newcommand{\bF}{\begin{figure}}
\newcommand{\eF}{\end{figure}}
\title{Gravity Induced New Topological Phase in Optics}
\author{Partha Ghose and M. K. Samal}
\address{S. N. Bose National Centre for Basic Sciences,
Block JD, Sector III, Salt Lake, Kolkata 700 098}
\begin{document}
\maketitle
\begin{abstract}

It is shown that both classical and quantum light can acquire a
topological phase shift induced by classical gravity, and the
latter is detectable in a laboratory-scale experiment.

\end{abstract}


Ever since the Aharonov-Bohm (AB) effect\cite{Aha} was
theoretically predicted, there has been the assumption that it is
a purely quantum mechanical effect without any classical
counterpart. The electrodynamic potentials themselves suddenly
acquired a physical significance that was lacking in classical
electrodynamics. However, Berry\cite{berry} has shown that the
Pancharatnam phase\cite{panch} for polarised classical light is
analogous to the AB phase. It is the purpose of this paper to show
that classical light can also exhibit a topological phase similar
to the AB phase and which is gravity-induced. This type of effect
holds in classical physics as long as (a) the classical system
under consideration has a wave-like character (as in
electrodynamics) and (b) the space is not simply connected. A very
important consequence of this classical optical analogue of the AB
effect implies that the classical gravitational `potential' has a
physical significance analogous to that of the electromagnetic
potentials in quantum mechanics. In fact, all potentials will
acquire physical significance in spaces that are not simply
connected.


To see the classical optical analogue of the scalar AB effect let
us consider a typical interference experiment with a Mach-Zehnder
interferometer, a pulsed source of classical light (with
Poissonian statistics) of energy $E$ (and therefore of effective
mass $E/c^2$) and a thin spherical shell of mass $ M $ and radius
$R$ placed in the path of one of the two beams inside the
interferometer. The spherical shell has two small holes
diametrically opposite each other that allow the light pulses to
pass through its field-free interior. The interaction Hamiltonian
$H_{int}$ inside the shell can be written as \be H_{int}= V_g
\frac{E}{c^2} = - \frac{G M}{R} \frac{E}{c^2} = -  \frac{G M}{R}
(\frac{4 \pi R^3 }{3 c^2} \frac{\epsilon_0 {\bf E} . {\bf E}}{2})
\equiv - \frac{4}{3} \pi R^3 . \frac{1}{2} \epsilon_g {\bf E} .
{\bf E} \ee where $\epsilon_g$ is the effective gravity-induced
permitivity seen by the light with electric vector ${\bf E}$
although it does not experience any gravitational force. It
follows from this that the effective refractive index is given by
\be
n = \sqrt{\epsilon_0 + \epsilon_g} \approx \sqrt{\epsilon_0} (1 +
\frac{G M}{2 R c^2})\ee since $\epsilon_g << \epsilon_0$.
Therefore light emerging from the shell will experience a
gravity-induced topological phase shift given by
\be
\phi_{cl} = \frac{2 \pi}{\lambda} (2 R) (n - \sqrt{\epsilon_0})
\approx \frac{2 \pi G M }{\lambda c^2} \sqrt{\epsilon_0}   \ee
where $\lambda$ is the wavelength of light.

The constant potential $V = GM/R$ cannot be gauged away because
the spaces inside and outside the sphere are not simply connected.
Therefore, this phase shift will be observable, and is the new
optical analogue of the scalar AB shift pointed out by Aharonov
and Bohm in their original paper.


The quantum optical analogue of the scalar AB effect can be
studied by writing the classical Maxwell theory of electrodynamics
in the Schr\"{o}dinger form\cite{Kemmer}

\be
i \hbar \frac{\partial (\gamma \psi)}{\partial t} = (H_0 +
H_{int}) (\gamma \psi)\label{eq:1}\ee with $\psi^T =
\frac{1}{\sqrt{m}}( -E_x, -E_y, -E_z, H_x, H_y, H_z, -m A_x, -m
A_y, -m A_z, m A_0)$ and

\be
H_0 = -i \hbar \tilde{\beta}_i \partial_i \psi \ee where
$\tilde{\beta}_i = \beta_0 \beta_i - \beta_i \beta_0$, the
$\beta$'s being the ($10\times10$) Kemmer-Duffin-Petiau matrices
satisfying the algebra

\be
\beta_{\mu} \beta_{\nu} \beta_{\lambda} + \beta_{\lambda}
\beta_{\nu} \beta_{\mu} = \beta_{\mu} g_{\nu \lambda} +
\beta_{\lambda} g_{\nu \mu} \ee and

\be
\gamma \beta_{\mu} + \beta_{\mu} \gamma = \beta_{\mu} \ee The
matrix $\gamma$ is diagonal with the first six diagonal elements
unity and the last four diagonal elements zero. Equation
(\ref{eq:1}) must be supplemented by the gauge-invariant
first-class constraint equation

\be
i\hbar \beta_i \beta_0^2 \partial_i \psi + m (1 - \beta_0^2)
\gamma \psi = 0\label{eq:2} \ee Equations (\ref{eq:1}) and
(\ref{eq:2}) together are equivalent to the Lorentz invariant
equation

\be
i \hbar \beta_{\mu} \partial^{\mu} \psi + m \gamma \psi = 0 \ee
which implies the massless second-order equation

\be
\Box(\gamma \psi) = 0 \label{eq:box} \ee The Schr\"odinger
equation (\ref{eq:1}) translates into the Maxwell equations

\ben \rm{curl}\, {\bf H} &=& \frac{1}{c}\frac{\partial {\bf
E}}{\partial t}\nonumber\\ \rm{curl}\, {\bf E} &=& -
\frac{1}{c}\frac{\partial {\bf H}}{\partial t}\label{eq:mx1} \een
and the constraint equation (\ref{eq:2}) into the equations

\ben \rm{div}\, {\bf E} &=& 0\nonumber\\ {\bf H} &=& \rm{curl}\,
{\bf A}\label{eq:mx2} \een This set of four equations together
constitute the complete set of Maxwell equations. This shows that
although the quantum mechanical equation (\ref{eq:1}) involves
$\hbar$, the Maxwell equations which are contained in it do not.
Hence it is not possible to say whether the Maxwell equations by
themselves describe a classical or quantum system. It is nowadays
well known that the quantum nature of radiation is difficult to
observe, and depends crucially on how light is produced. In most
production processes the light produced has Poissonian statistics
and is indistinguishable from classical light. It is only when
special care is taken to produce Fock states of light (as in
parametric down-conversion of laser light) or sqeezed states that
the quantum nature of radiation can be observed. {\it Hence the
production process constitutes the additional constraint that
renders the Maxwell equations either classical or quantum.}

There are three features of this formalism that must be pointed
out. First, there is a conserved four-vector current $s_{\mu}$ in
this theory whose time component $s_o$ is positive definite, and
can be identified with the probability current\cite{Ghose2}. It is
associated with the energy flux and energy density. This makes a
consistent quantum mechanical interpretation of this formalism
possible. Second, the formalism describes a massless wave/particle
in spite of the mass parameter $m$ which is essential for
dimensional reasons. Nevertheless, the theory is gauge invariant
under the local transformations $\psi \rightarrow \psi + (1 -
\gamma)\chi$ where $\chi$ is an arbitrary function of space and
time. In fact, this mass parameter can be expressed in terms of
the fundamental constants of the theory by the relation $m = s
\hbar \omega/c^2$ where $\omega$ is the frequency of the photon
and $s = 2$ is the spin degeneracy. This aspect of the formalism
has not been pointed out before. Third, since the classical
Maxwell equations (\ref{eq:mx1}) and (\ref{eq:mx2}) are contained
in this quantum mechanical formalism but do not involve any
dimensional mass parameter, it is clear that {\it the classical
angular momentum as well as the classical polarization states are
totally independent of the quantum mechanical spin $\sigma$}. We
are not aware of any other formalism in which this distinction
between the quantum mechanical spin of the photon and the angular
momentum of the classical electromagnetic field is so clearly
expressed.

If one considers the same type of experiment discussed in the
previous section but with a pulse of laser light of energy $E
\approx \bar{N} h \nu$ where $\bar{N}$ is the average number of
photons in the pulse around which the distribution is peaked, then
the phase shift predicted by the Schr\"odinger equation
(\ref{eq:1}) is
\be
\phi_{qm} = \frac{H_{int} t}{\hbar} = \frac{4 \pi G M
\bar{N}}{\lambda c^2}  \ee since $\epsilon_0 = 1$. The advantage
of using a laser pulse is that the phase shift is enhanced by the
factor $\bar{N}$ compared to a single photon state. Using a
typical pico second pulsed laser with wavelength $\approx 5000
{\rm \AA} $ one gets $\bar{N} \approx 10^7$. This enhancement is
not possible in the classical limit which corresponds to $\bar{N}
<< 1$. This can also be seen from the fact that an increase in the
intensity of classical light does not enhance the refractive index
(hence the phase shift) because of the linearity of the vacuum.


The question is whether the predicted phase shifts can be
detected.  For classical light of wavelength $\approx 5000 {\rm
\AA} $ one needs a thin spherical shell of mass $\approx 10^{18}$
kg to produce a phase shift of the order of a milliradian. This
kind of mass can only be realized in astrophysical conditions.
Nevertheless a laboratory-scale experiment could still be possible
by compensating for the large mass through an increase of the
effective optical path (through the sphere) by repeatedly
circulating the same light pulse as shown in the figure. This is
equivalent to introducing a large winding number $n_w$.

\begin{figure}[htbp]
\centerline{\epsfxsize = 0.75 \textwidth \epsfbox{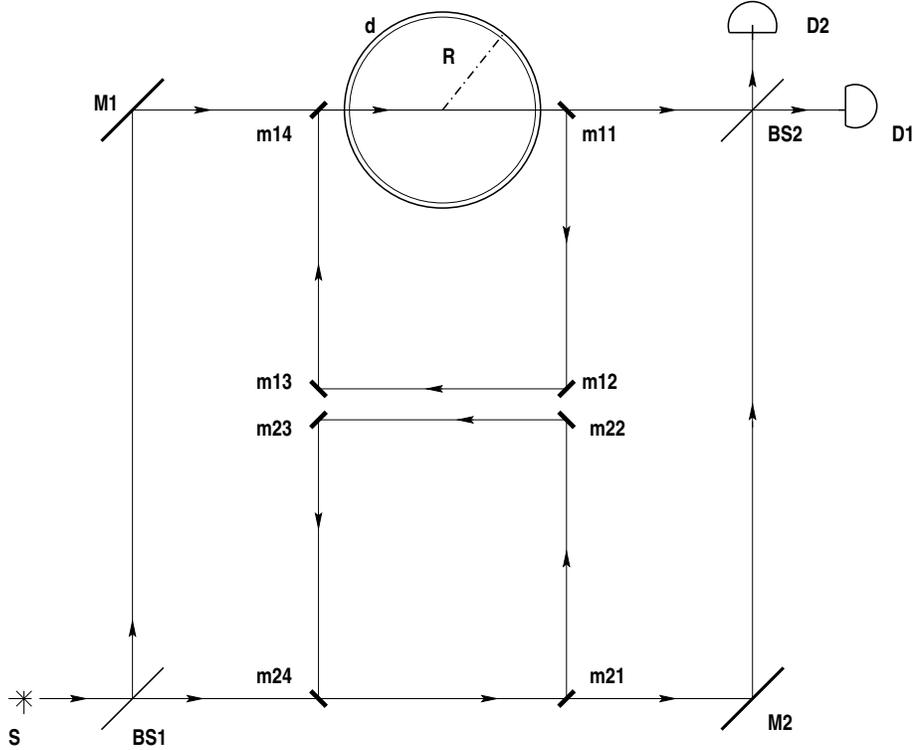}}
\caption{A modified Mach-Zehnder set-up }
\end{figure}

The figure shows a modified Mach-Zehnder set-up having additional
lossless mirrors {\sf mij} with {\sf m12, m13, m22, m23} fixed and
{\sf m11, m14, m21, m24} movable. {\sf m11} and {\sf m21} are
inserted from the beginning and removed after $n$ cycles are
completed by the light pulses. {\sf m14} and {\sf m24} are
inserted after the light pulses enter the sphere and in time for
them to circulate as shown.

Note that the light pulses acquire a dynamical phase while
circulating in the gravitational field between the mirrors. If the
paths between the fixed mirrors {\sf m12, m22} and {\sf m13, m23}
are kept very close, the dynamical phases picked up along these
paths will be nearly the same, and hence will cancel one another
in the interference pattern. The only additional dynamical phase
in the lower circuit will come from the path between the mirrors
{\sf m24} and {\sf m21}. This can be made significantly small
compared to the topological phase acquired in the upper circuit by
arranging the distance between the mirror {\sf M1} and the
beam-splitter {\sf BS1} large enough compared to the radius $R$ of
the spherical shell.

A simple calculation shows that for a thin spherical shell of
radius $\approx 3.3$ m, thickness $10$ cm and mass $\approx
10^{5}$ kg one needs a winding number $n_w = 10^{12}$ which would
require a duration of the order of $15$ hours. The problem in
actually performing the suggested experiment would be maintaining
the required stability of the entire set-up as well as the
lossless character of the mirrors to detect the phase-shift of the
order of a milliradian over such a long duration.

As we have seen, unlike the classical situation, the use of a
pulsed laser has the advantage of enhancing the phase shift by a
factor of $\bar{N} \approx 10^7$. This enables a reduction in the
dimensions of the spherical shell to radius $\approx 1.5$ m,
thickness $1$ cm and mass $\approx 3 \times 10^{3}$ kg for a
winding number $n_w = 10^{6}$. This can be achieved in a time
interval of the order of $0.1$ sec which makes the experiment
feasible.


The new result of this work ,is that both {\it classical} and
quantum light can acquire a topological phase that is induced by
{\it classical} gravity and the latter can be measured in a
laboratory-scale experiment. One important implication of this is
that the classical gravitational potential acquires a new physical
significance in spaces that are not simply connected.

We thank DST, Govt. of India for financial support to carry out
this work and A. Greentree for helpful correspondence.

\end{document}